\documentclass[%
reprint,
superscriptaddress,
nofootinbib,
 amsmath,amssymb,
 aps,
 prl
]{revtex4-2}

\usepackage{graphicx}
\usepackage{amsmath}
\usepackage{dcolumn}
\usepackage{bm}
\usepackage{hyperref}
\usepackage{xurl}
\usepackage{multirow}
\usepackage[mathlines]{lineno}
\usepackage{color}
\usepackage{ulem}
\usepackage{relsize}
\def\babar{\mbox{\slshape B\kern-0.1em{\smaller A}\kern-0.1em
    B\kern-0.1em{\smaller A\kern-0.2em R}}}
\usepackage[english]{babel}
\newcommand{\BESIIIorcid}[1]{\href{https://orcid.org/#1}{\hspace*{0.1em}\raisebox{-0.45ex}{\includegraphics[width=1em]{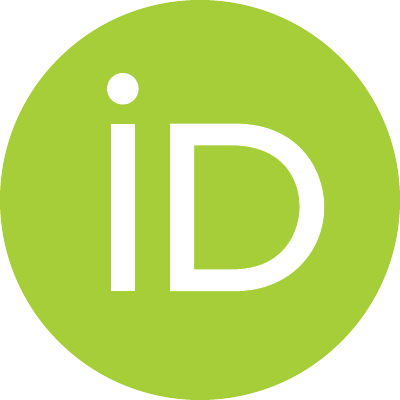}}}} 

\begin{document}

\preprint{APS/123-QED}

\title{Observation of double $s\bar{s}$ production in $e^+e^-$ collision at $\sqrt{s} = 3.08~\textrm{GeV}$}

\author{
  \small
M.~Ablikim$^{1}$\BESIIIorcid{0000-0002-3935-619X},
M.~N.~Achasov$^{4,c}$\BESIIIorcid{0000-0002-9400-8622},
P.~Adlarson$^{84}$\BESIIIorcid{0000-0001-6280-3851},
X.~C.~Ai$^{90}$\BESIIIorcid{0000-0003-3856-2415},
C.~S.~Akondi$^{32A,32B}$\BESIIIorcid{0000-0001-6303-5217},
R.~Aliberti$^{40}$\BESIIIorcid{0000-0003-3500-4012},
A.~Amoroso$^{83A,83C}$\BESIIIorcid{0000-0002-3095-8610},
Q.~An$^{79,66,\dagger}$,
Y.~H.~An$^{90}$\BESIIIorcid{0009-0008-3419-0849},
M.~S.~Anderson$^{40}$\BESIIIorcid{0009-0008-1550-2632}
% ...
\\
\vspace{0.2cm}
(BESIII Collaboration)\\
\vspace{0.2cm} {\it
$^{1}$ Institute of High Energy Physics, Beijing 100049, People's Republic of China\\
% ...
$^{4}$ Budker Institute of Nuclear Physics SB RAS (BINP), Novosibirsk 630090, Russia\\
% ...
$^{32}$ INFN Sezione di Ferrara, (A)INFN Sezione di Ferrara, I-44122, Ferrara, Italy; (B)University of Ferrara, I-44122, Ferrara, Italy\\
% ...
$^{40}$ Johannes Gutenberg University of Mainz, Johann-Joachim-Becher-Weg 45, D-55099 Mainz, Germany\\
% ...
$^{66}$ State Key Laboratory of Particle Detection and Electronics, Beijing 100049, Hefei 230026, People's Republic of China\\
% ...
$^{79}$ University of Science and Technology of China, Hefei 230026, People's Republic of China\\
% ...
$^{83}$ University of Turin and INFN, (A)University of Turin, I-10125, Turin, Italy; (B)University of Eastern Piedmont, I-15121, Alessandria, Italy; (C)INFN, I-10125, Turin, Italy\\
$^{84}$ Uppsala University, Box 516, SE-75120 Uppsala, Sweden\\
% ...
$^{90}$ Zhengzhou University, Zhengzhou 450001, People's Republic of China\\
\vspace{0.2cm}
$^{\dagger}$ Deceased\\
% ...
$^{c}$ Also at the Novosibirsk State University, Novosibirsk, 630090, Russia\\
% ...
}
}


\date{\today}

\begin{abstract}

We report the observation of significant double-$s\bar{s}$ production in the $e^+e^-$ continuum, based on the measurement of prompt $\phi$ mesons produced in association with hadrons containing an $s$ quark or an $s\bar{s}$ pair. In an analysis of $e^+e^-$ collision data collected by the BESIII experiment at $\sqrt{s}=3.08~\textrm{GeV}$, the ratio $\sigma(e^+e^- \to \phi\ s\bar{s}+\textrm{anything}) / \sigma(e^+e^-\rightarrow\phi+\textrm{anything})$ is determined to be $(40.4\pm1.7_{\rm stat.}\pm1.5_{\rm syst.})\%$ by detecting and measuring $e^+e^-\to\phi + X(s\bar{s})$,
where $X(s\bar{s})$ denotes an $\eta$ meson, an $\eta^{\prime}$ meson, or one of the strange-meson pairs $K^+K^-$, $K^+K^{*-}$, $K^-K^{*+}$, $K^0\bar{K}^{0}$, and $K^0\bar{K}^{*0}+\textrm{c.c.}$.
The level of double-$s\bar{s}$ production is in line with the double-$c\bar{c}$ production reported by the Belle and \babar\ collaborations, for which theoretical calculations predict lower rates. The experimental measurement of double $s\bar{s}$ production at BESIII can shed light on the understanding of quark hadronization and QCD.

\end{abstract}

\maketitle

The production of hadrons in $e^+e^-$ annihilation provides a unique and clean probe of the transition from quarks $q$, antiquarks $\bar{q}$, and gluons $g$ to color-neutral hadrons, which is a process fundamental to quantum chromodynamics (QCD) and crucial for the formation of visible matter.
Furthermore, experimental investigation of possible correlations in hadronization between different quark pairs offers access to studying the properties of QCD.
The Belle~\cite{ref:Belle_2002, ref:Belle_2004abn, Belle:2009bxr} and \babar\ Collaborations~\cite{ref:BaBar_2005} observed significant double-$c\bar{c}$ production, with the process $e^+e^-\rightarrow c\bar{c}c\bar{c} + \textrm{anything}$, using $e^+e^-$ collision data at a center-of-mass energy ($\sqrt{s}$) of $10.6~\text{GeV}$.
In double-$c\bar{c}$ production,
one $c\bar{c}$ pair hadronizes into a $J/\psi$ meson, while the accompanying pair yields either a second charmonium state or open-charm hadrons via fragmentation.
The measurements of double-$c\bar{c}$ production revealed a significant excess over theoretical predictions in the ratio $R_{c\bar{c}} = \sigma(e^+e^-\rightarrow J/\psi\ c\bar{c} + \textrm{anything})/\sigma(e^+e^-\rightarrow J/\psi + \textrm{anything})$, 
which prompted several efforts to address the discrepancy and advance the understanding of QCD~\cite{Braaten:2002fi, Bodwin:2002fk, Liu:2002wq, Hagiwara:2003cw, Liu:2004ga, Ma:2008gq}.
Although subsequent theoretical refinements have significantly alleviated the discrepancy in $R_{c\bar{c}}$ between QCD predictions and experimental data,
a large tension still remains in the cross sections for several double $c\bar{c}$ production processes~\cite{Braaten:2002fi, Liu:2002wq, Hagiwara:2003cw, Liu:2004ga}.
This phenomenon suggests unforeseen dynamics in hadronization, which are essential for a complete theoretical understanding of QCD.

The double-$s\bar{s}$ production process, $e^+e^-\rightarrow s\bar{s} s\bar{s} + \textrm{anything}$,
is a critical testing ground for phenomena analogous to those observed in double-$c\bar{c}$ production.
Produced at $\sqrt{s}=3.08$~GeV, double-$s\bar{s}$ provides an ideal laboratory for comprehensive studies of QCD in the non-perturbative regime. Furthermore, the synergy of the strange and charm flavor portals provides a direct experimental window to explore the possible flavor correlation in hadronization.

Previous individual measurements related to double-$s\bar{s}$ production have been performed using $e^+e^-$ collision data, such as $e^+e^-\rightarrow\phi\eta / \eta^{\prime} / K^+K^-$~\cite{BESIII:2021bjn, BESIII:2020gnc, BESIII:2019ebn}.
However, the absence of systematic measurements in the study of double-$s\bar{s}$ production indicates a critical gap in understanding the observed phenomena in double-$c\bar{c}$ production and the underlying hadronization mechanisms. Such systematic measurements are therefore of high interest.

In this Letter, we present a systematic study of the $e^+e^-\rightarrow s\bar{s}s\bar{s} + \textrm{anything}$ process by analyzing $167.06~\textrm{pb}^{-1}$~\cite{BESIII:2016kpv, BESIII:2021cxx, BESIII:2017lkp} of $e^+e^-$ collision data collected at $\sqrt{s} = 3.08~\textrm{GeV}$ with the BESIII detector~\cite{BESIII:2009fln},
where one $s\bar{s}$ pair hadronizes into a $\phi$ meson, and the other pair produces additional hadrons that carry the remaining strangeness, including
$\eta$, $\eta^{\prime}$, $K^+K^-$, $K^+K^{*-}$, $K^-K^{*+}$, $K^0\bar{K}^{0}$, and $K^0\bar{K}^{*0} + \textrm{c.c.}$.
The cross sections of these double-$s\bar{s}$ production processes and $e^+e^-\rightarrow \phi + \textrm{anything}$ are measured in this Letter.

A ratio, $R_{s\bar{s}}$, is calculated from the measured cross sections:
\begin{equation}
    R_{s\bar{s}} = \frac{\sum \sigma(e^+e^-\rightarrow\phi + X(s\bar{s})) f_{s\bar{s}}}{\sigma(e^+e^-\rightarrow\phi + \textrm{anything})},
    \label{eq:Rss}
\end{equation}
where $X(s\bar{s})$ represents the $\eta$, $\eta^{\prime}$, $K^+K^-$, $K^+K^{*-}$, $K^-K^{*+}$, $K^0\bar{K}^{0}$, and $K^0\bar{K}^{*0} + \textrm{c.c.}$ states, and $f_{s\bar{s}}$ is the fraction of the $s\bar{s}$ component in $X(s\bar{s})$.

The BESIII detector records $e^+e^-$ collisions provided by the BEPCII storage ring~\cite{Yu:2016cof} in the $\sqrt{s}$ range from 1.84 to $4.95~\text{GeV}$, 
with a peak luminosity of $1.1 \times 10^{33}~\textrm{cm}^{-2}\textrm{s}^{-1}$ achieved at $\sqrt{s} = 3.773~\textrm{GeV}$.
The cylindrical core of the BESIII detector covers 93\% of the full solid angle and consists of a helium-based multi-layer drift chamber~(MDC), a plastic scintillator time-of-flight~(TOF) system, and a CsI(Tl) electromagnetic calorimeter~(EMC), all enclosed in a superconducting solenoidal magnet providing a 1.0~T magnetic field.
The solenoid is supported by an octagonal flux-return yoke with resistive plate counter muon identification modules interleaved with steel.
The charged-particle momentum resolution at $1~\text{GeV}/c$ is $0.5\%$, and the ${\rm d}E/{\rm d}x$ resolution is $6\%$ for electrons from Bhabha scattering. 
The EMC measures photon energies with a resolution of $2.5\%$ ($5\%$) at $1$~GeV in the barrel (end cap) region. 
The time resolution in the TOF barrel region is 68~ps, while that in the end cap region is 110~ps. 
The end cap TOF system was upgraded in 2015 using multigap resistive plate chamber technology, providing a time resolution of 60~ps~\cite{li2017study, guo2017study, cao2020design}.

Simulated data samples produced with a {\sc geant4}-based~\cite{GEANT4:2002zbu} Monte Carlo~(MC) package, which includes the geometric description~\cite{Huang:2022wuo} of the BESIII detector and its response, are used to determine detection efficiencies and to estimate background contributions.
Initial-state radiation~(ISR) in $e^+e^-$ annihilations is modeled with {\sc ConExc}~\cite{Ping:2013jka}.
All particle decays are modeled with {\sc evtgen}~\cite{Lange:2001uf, Ping:2008zz} using branching fractions taken from the Particle Data Group~\cite{ParticleDataGroup:2022pth} when available,
or estimated using {\sc lundcharm}~\cite{Chen:2000tv, Yang:2014vra}.
Final-state radiation from charged final-state particles is incorporated using the {\sc photos} package~\cite{Richter-Was:1992hxq}.
Exclusive MC samples are generated for the double-$s\bar{s}$ production processes,
where experimental knowledge is implemented for $e^+e^- \to \phi\eta/\eta'$~\cite{BESIII:2021bjn, BESIII:2020gnc}.
The Hybrid generator developed for inclusive hadron processes~\cite{Ping:2016pms} is employed to generate an inclusive MC sample.

We select events of interest by requiring at least two oppositely charged tracks, and additionally one or two charged tracks as required.
Charged tracks detected in the MDC are required to satisfy $|\cos\theta| < 0.93$, where $\theta$ denotes the polar angle with respect to the $z$-axis, the symmetry axis of the MDC.
The distance of closest approach to the interaction point must be less than 10~cm along the $z$-axis and less than 1~cm in the transverse plane,
except for the charged tracks used to reconstruct $K_S^0$ candidates.
Particle identification (PID) for charged tracks integrates measurements of the ionization energy loss in the MDC (d$E$/d$x$), the flight time from the TOF system, and the shower energy deposited in the EMC to formulate likelihoods $\mathcal{L}(h)~(h=p,K,\pi)$ for each hadron hypothesis $h$. 
Each track is identified as the particle type corresponding to the most probable hypothesis. 
The $K_S^0$ candidate is reconstructed using two charged tracks consistent with the pion PID hypothesis and with opposite charges.
To reduce background contributions, $\pi^+\pi^-$ pairs are constrained to originate from a common vertex.
The distance between this common vertex and the primary vertex is defined as the decay length of the $K_S^0$, and the significance of this decay length is required to be larger than two standard deviations. 

\begin{figure}[!htbp]
    \centering
    \includegraphics[scale=0.4]{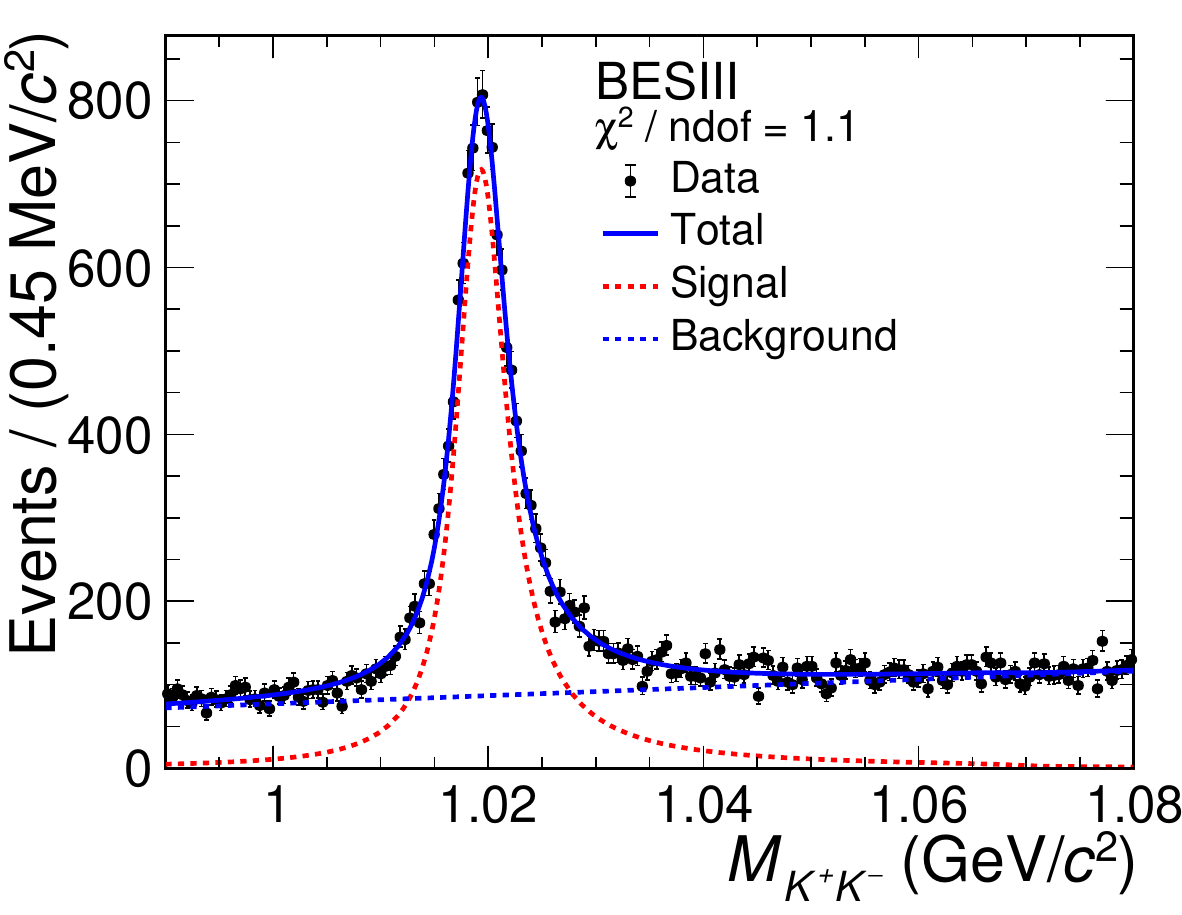}
    \caption{The fit result for $M_{K^+K^-}$. The points with error bars are data. The solid blue line, dashed red line, and dashed blue line represent the total fit, the $e^+e^-\rightarrow\phi + \textrm{anything}$, and the background, respectively.}
    \label{fig:phi}
\end{figure}

Two charged tracks with opposite charges and consistent with the kaon PID hypothesis are required to reconstruct the $\phi$ candidate.
If there are multiple $K^+K^-$ combinations within a single event,
all combinations are retained for the analysis. 
The invariant mass of the $K^+K^-$ combination, 
$M_{K^+K^-} = \sqrt{(p_{K^+} + p_{K^-})^{2}}$, 
is used to determine the yields of $e^+e^-\rightarrow\phi + \textrm{anything}$,
where $p$ is the four-momentum of the corresponding particle candidate.
A fit to the $M_{K^+K^-}$ spectrum is performed to extract the yield of $e^+e^-\rightarrow\phi + \textrm{anything}$ events.
In the fit, a Breit--Wigner function~\cite{ParticleDataGroup:2022pth} convolved with a double-sided Crystal Ball function~\cite{Skwarnicki:1986xj} is taken as the signal model, while a first-order Chebyshev polynomial is used as the background model.
The fit result is shown in Fig.~\ref{fig:phi}.

\begin{figure}[!htbp]
    \centering
    \includegraphics[scale=0.4]{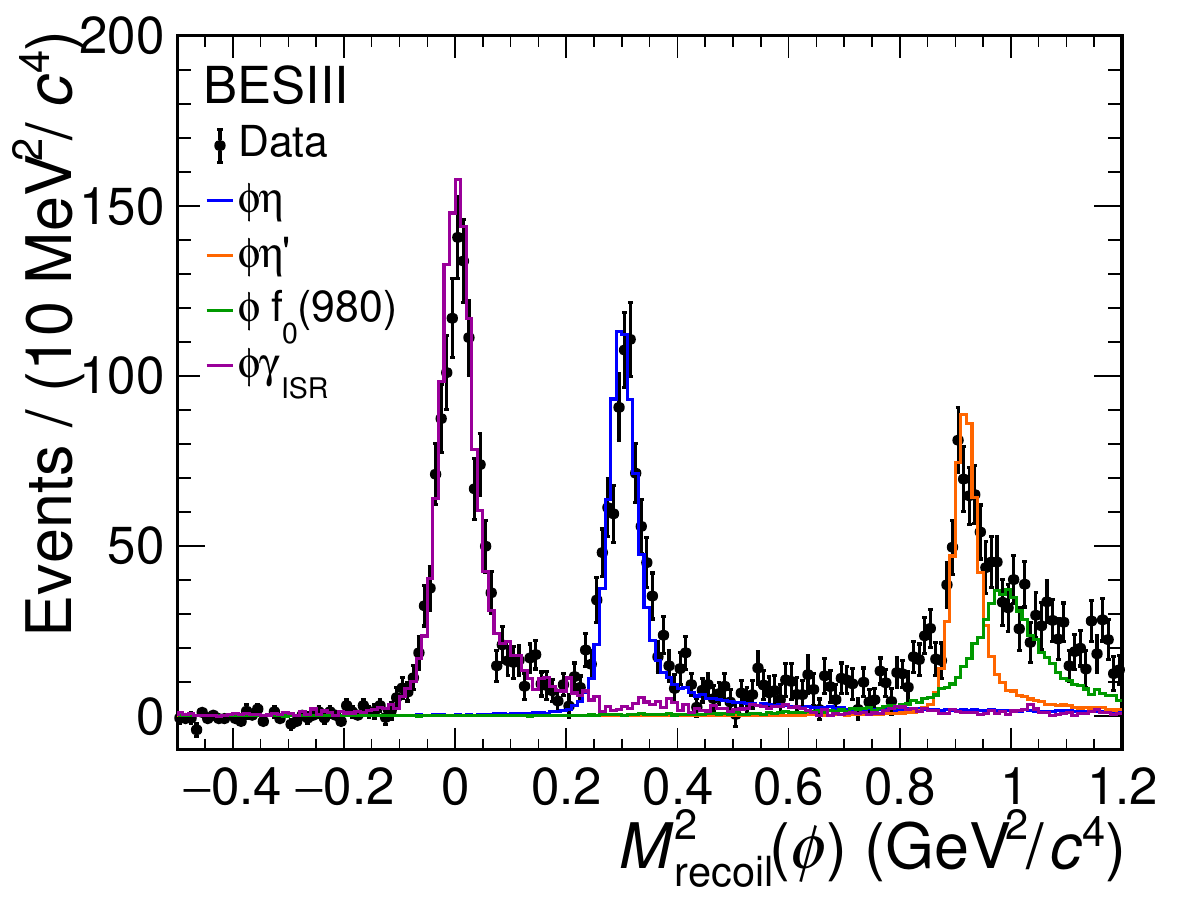}
    \caption{The distribution of $M^2_{\textrm{recoil}}(\phi)$ after non-$\phi$ background subtraction. The points with error bars are data, while the lines are MC distributions: the blue, orange, green, and purple lines represent $e^+e^-\to\phi\eta$, $e^+e^-\to\phi\eta^{\prime}$, $e^+e^-\to\phi f_0(980)$, $e^+e^-\to\phi\gamma_{\textrm{ISR}}$, respectively. The MC samples are normalized arbitrarily.}
    \label{fig:recoil}
\end{figure}

To further select double-strange events containing a $\phi$ meson, the requirement $1.010 < M_{K^+K^-} < 1.030~\textrm{GeV}/c^2$ is applied.
The spectrum of recoil mass against the $\phi$ candidate, 
$M_{\textrm{recoil}}(\phi) = \sqrt{(p_{e^+e^-} - p_{\phi})^{2}}$, 
is used to determine the cross sections of the double-$s\bar{s}$ production processes and $R_{s\bar{s}}$,
where $p_{e^+e^-}$ is the four-momentum of the initial $e^+e^-$ system~\cite{BESIII:2021cxx}.
After the above selection, the dominant background events originate from processes without explicit $\phi$ production, as estimated by the inclusive MC sample and referred to as non-$\phi$ background.
These background candidates can be estimated using the $\phi$ sideband regions in data, which are chosen as $0.990 < M_{K^+K^-} < 0.995~\text{GeV}/c^{2}$ and $1.043 < M_{K^+K^-} < 1.060~\text{GeV}/c^{2}$, based on the fit results shown in Fig.~\ref{fig:phi}.
After subtracting these background candidates, the distribution of $M^2_{\textrm{recoil}}(\phi)$ is shown in Fig.~\ref{fig:recoil}.
Significant peaks corresponding to the $\eta$ and $\eta^{\prime}$ mesons are observed in the region below $4 m_{K}^{2}$, where $m_{K}$ is the kaon mass. These mesons constitute part of the double-$s\bar{s}$ production signal.

Among the event candidates with a $\phi$ meson,
an additional charged track identified as a kaon is required to select events of the $e^+ e^- \to \phi K^+ / K^- + \textrm{anything}$ processes,
while at least one $K_S^0$ is required to reconstruct the $e^+e^- \to \phi K^0 + \textrm{anything} +\textrm{c.c.}$ processes, and if there are multiple $K_S^0$ candidates within a single event, all the $K_S^0$s are retained for the analysis.
A two-dimensional sideband, defined in terms of $M_{K^+K^-}$ and the invariant mass of the $\pi^{+}\pi^{-}$ combination, is used to further suppress the background events in the $e^+e^-\to\phi K^{0} \textrm{anything} +\textrm{c.c.}$ process.

The recoil masses against the $\phi K^+/K^-/K^0_S$, 
\begin{equation}
    \begin{aligned}
        & M_{\textrm{recoil}}(\phi K^+ / K^- / K^0_S) =  \\
        & \sqrt{(p_{e^+e^-} - p_{\phi} - p'_{K^+}/p'_{K^-}/p_{K^0_S})^{2}}, \\
    \end{aligned}
    \label{eq:recoil_mass}
\end{equation}
are calculated to determine the yields of the $e^+e^- \to \phi K^+K^- (e^+e^- \to \phi K^+K^{*-}+\textrm{c.c.})$ and $e^+e^- \to \phi K^0\bar{K}^0 (e^+e^- \to \phi K^0\bar{K}^{*0}+\textrm{c.c.})$ events, respectively, where $p'_{K^+}$, $p'_{K^-}$ and $p_{K^0_S}$ are the four-momenta of the additional $K^+$, additional $K^-$, and $K^0_S$ candidates, respectively.

\begin{figure*}[!htbp]
\centering
\includegraphics[scale=0.43]{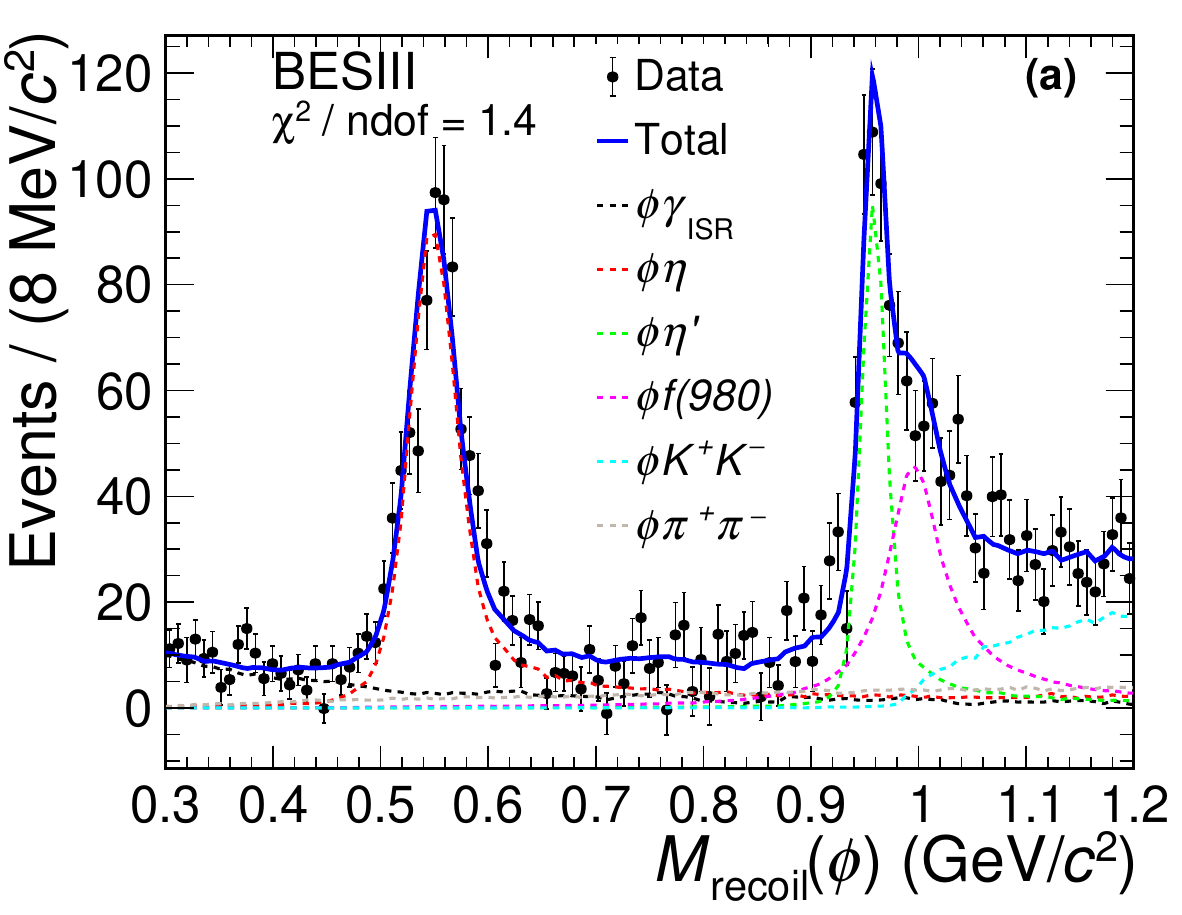}
\includegraphics[scale=0.43]{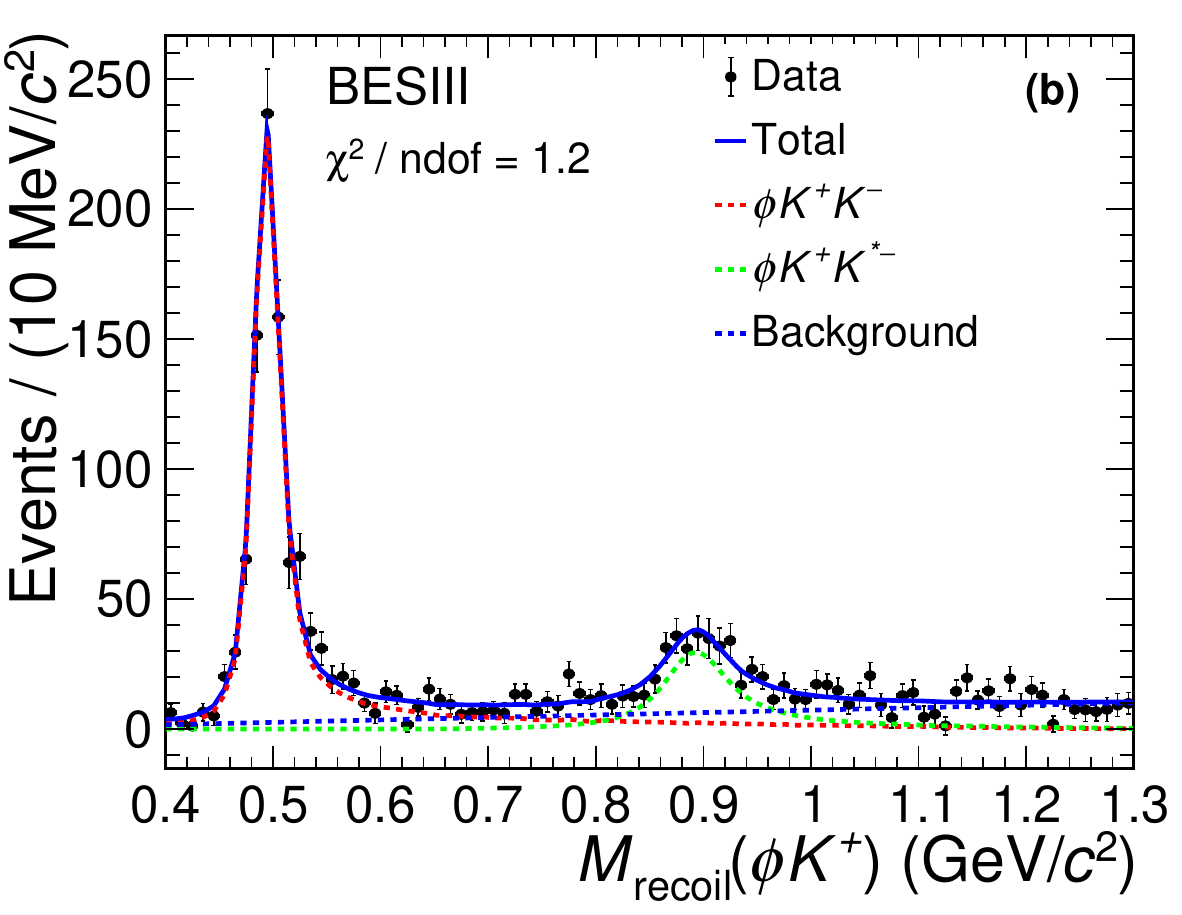}
\includegraphics[scale=0.43]{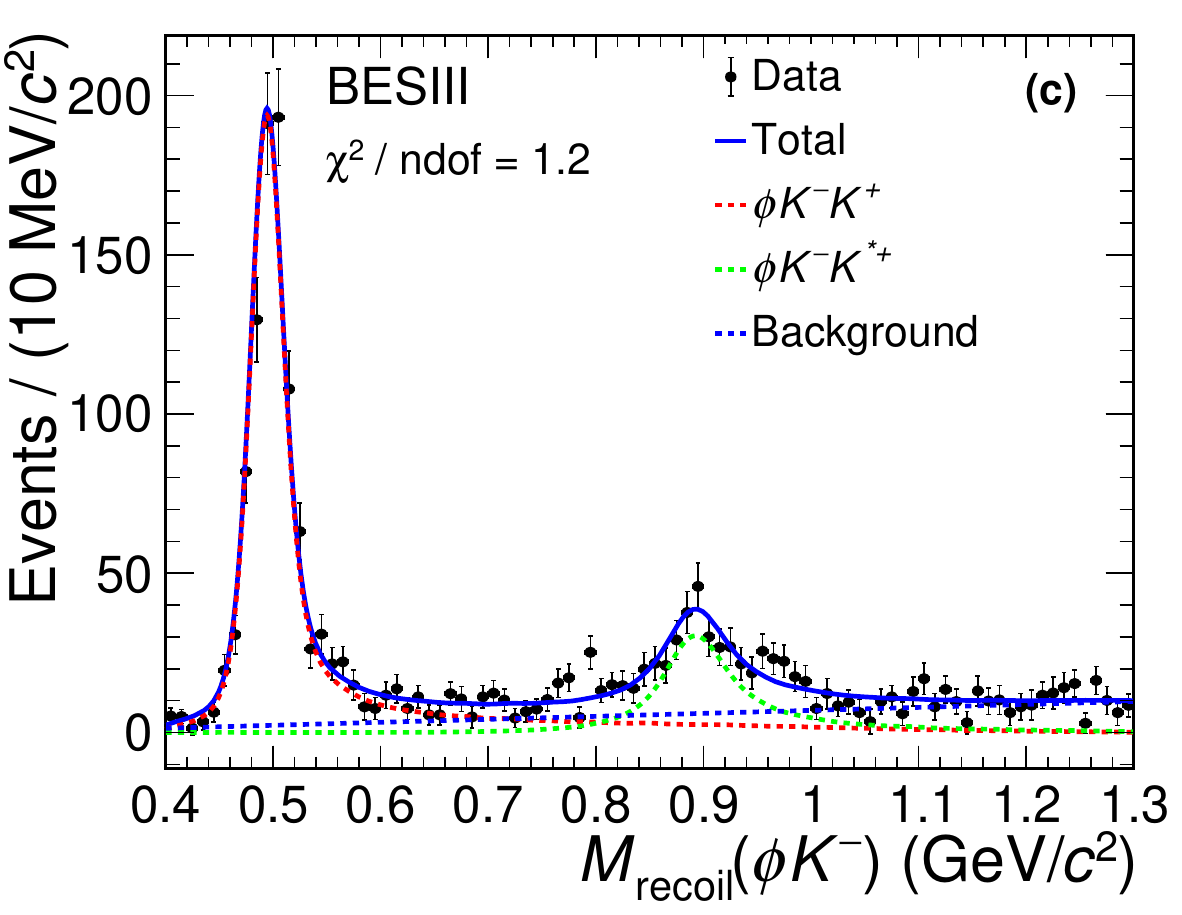}
\includegraphics[scale=0.43]{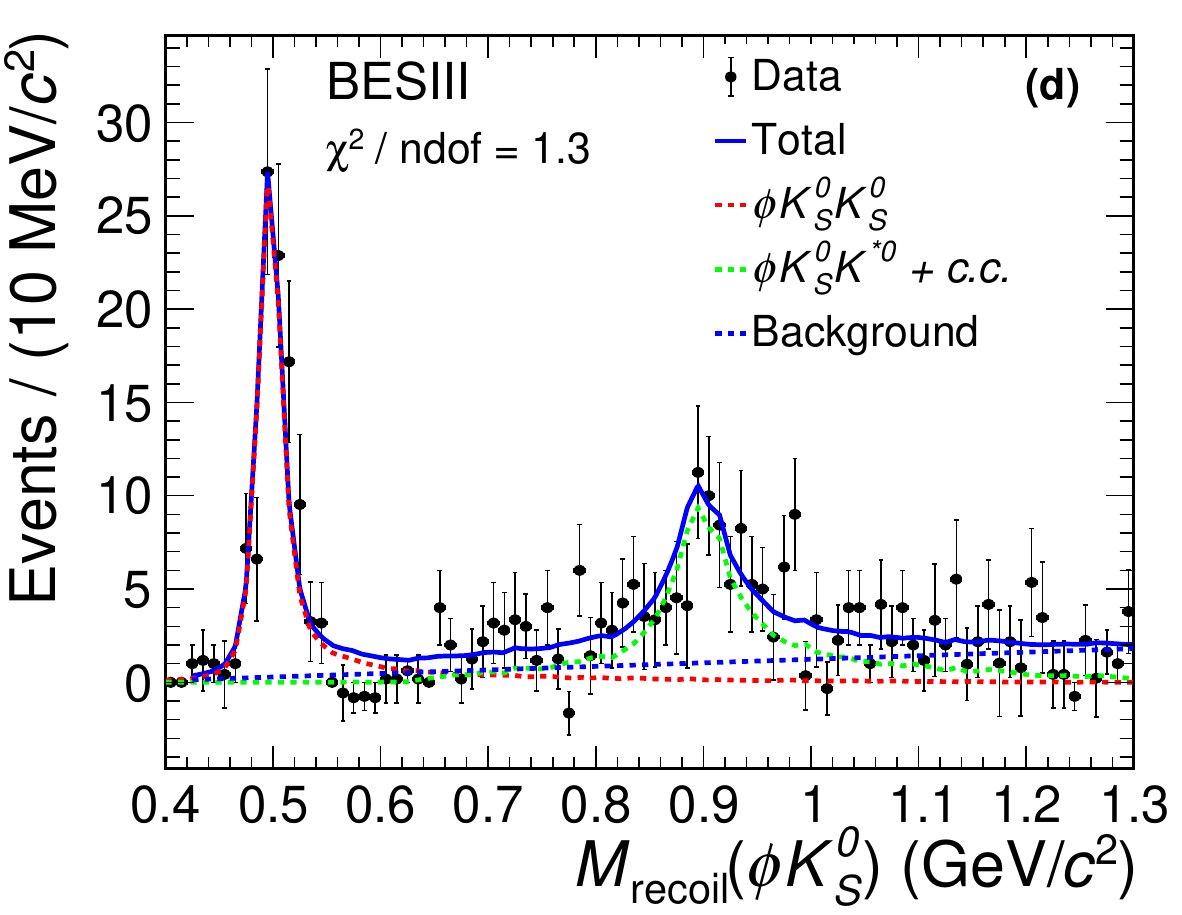}
\caption{
  The fit results for (a) $M_{\textrm{recoil}}(\phi)$, (b) $M_{\textrm{recoil}}(\phi K^{+})$, (c) $M_{\textrm{recoil}}(\phi K^{-})$, and (d) $M_{\textrm{recoil}}(\phi K^0_{S})$.
  In each plot, the points with error bars are data, and the solid blue line is the total fit.
  In (a), the dashed lines represent MC sample shapes: the ISR, $\eta$, $\eta'$, $f_{0}(980)$, $K^+K^-$, and $\pi^+\pi^-$ components are shown as black, red, green, magenta, light blue, and gray dashed lines, respectively.
  In (b), (c), and (d), the dashed blue line is the background; 
  the dashed red line corresponds to $e^+e^-\rightarrow\phi K^+K^-$ for $M_{\textrm{recoil}}(\phi K^{\pm})$, and $e^+e^-\rightarrow\phi K^0_LK^0_{S/L}$ for $M_{\textrm{recoil}}(\phi K^0_S)$;
  and the dashed green line is $e^+e^-\rightarrow\phi K^+K^{*-}$ for $M_{\textrm{recoil}}(\phi K^+)$, $e^+e^-\rightarrow\phi K^-K^{*+}$ for $M_{\textrm{recoil}}(\phi K^-)$, and $e^+e^-\rightarrow\phi K^0_SK^{*0}/\bar{K}^{*0}$ for $M_{\textrm{recoil}}(\phi K^0_S)$.
}
\label{fig:fit_recoil}
\end{figure*}

The observed cross sections of $e^+e^-\to\phi+\text{anything}$ and $e^+e^-\to\phi+X(s\bar{s})$ are calculated via
\begin{equation}
    \sigma_i = \frac{n_i}{\mathcal{L}\times\epsilon_i \times\mathcal{B} (\phi\to K^+K^-)},
    \label{eq:xs}
\end{equation}
where $\mathcal{L}$ is the integrated luminosity of the $e^+e^-$ collision data at $\sqrt{s}=3.08~\text{GeV}$,
$n_i$ and $\epsilon_i$ are the yield and the selection efficiency for process $i$,
and $\mathcal{B} (\phi\to K^+K^-) = (49.9\pm0.5)\%$ is the branching fraction of $\phi\to K^+K^-$~\cite{ParticleDataGroup:2022pth}.
The yields for the $e^+e^-\to\phi+X(s\bar{s})$ processes are extracted from fits to the mass spectra of $M_{\textrm{recoil}}(\phi)$, $M_{\textrm{recoil}}(\phi K^+)$, $M_{\textrm{recoil}}(\phi K^-)$, and $M_{\textrm{recoil}}(\phi K_S^0)$, respectively.
In the fit to the $M_{\textrm{recoil}}(\phi)$ distribution, the signal shapes of $\phi\eta$, $\phi\eta^{\prime}$, $\phi\gamma_{\textrm{ISR}}$, $\phi\pi^+\pi^-$, $\phi K^+K^-$, and $\phi f_0(980)$ are obtained from the corresponding MC samples, while the yields of these components are left free.
In the fits to $ M_{\textrm{recoil}}(\phi K^+ / K^-)$,
the signal line shapes of $\phi K^+ K^-$, $\phi K^+ K^{*-}$, and $\phi K^- K^{*+}$ are taken from simulated samples convolved with Gaussian functions,
while the background shapes are parameterized by three different linear functions.
Additionally, the signal line shapes of $\phi K^0 \bar{K}^0$ and $\phi K^0 \bar{K}^{*0}$ are modeled using the corresponding MC samples,
and a first-order Chebyshev polynomial is used for the background shape.
The fit results for $M_{\textrm{recoil}}(\phi)$, $M_{\textrm{recoil}}(\phi K^+)$, $M_{\textrm{recoil}}(\phi K^-)$, and $M_{\textrm{recoil}}(\phi K_S^0)$ are shown in Fig.~\ref{fig:fit_recoil}.
The event yields for $e^+e^-\to\phi+\text{anything}$ and $e^+e^-\to\phi+X(s\bar{s})$ from the fit are summarized in Table~\ref{tab:xs_fit_res}.
The selection efficiency for $e^+e^-\rightarrow\phi + \textrm{anything}$ is estimated by reweighting the differential selection efficiency from the inclusive MC sample as a function of $\phi$ momentum.

The efficiencies for the individual $e^+e^-\to\phi+X(s\bar{s})$ processes are taken from the corresponding MC samples, as listed in Table~\ref{tab:xs_fit_res}. 
Neglecting CP violation and accounting for the fact that $K_L^0$ mesons are not reconstructed in BESIII, $\sigma(e^+e^-\rightarrow\phi K^0 \bar{K}^{0})$ is expected to be 2 times of $\sigma(e^+e^-\rightarrow\phi K_S^0 K_S^0)$.
\begin{table}[!htbp]
    \caption{The selection efficiency $\epsilon_i$, the fitted number of events $n_i$, and the cross section $\sigma_i$ are listed for each process $i$. The uncertainties on $\epsilon_i$ and $n_i$ are statistical only; for $\sigma_i$, the first uncertainty is statistical and the second is systematic.}
    \centering
    \begin{tabular}{lccc}
    \hline
    \hline
    Process                                   & $\epsilon_i$(\%) & $n_i$           & $\sigma_i(\textrm{pb})$ \\ \hline
    $\phi + \rm anything$                     & 27.5 $\pm$ 0.4   & 11618 $\pm$ 158 & 506.3$\pm$6.9$\pm$19.7  \\
    $\phi   \eta$                             & 28.5 $\pm$ 0.1   & 924 $\pm$ 39    & 38.9$\pm$1.6$\pm$1.6    \\
    $\phi   \eta^{\prime}$                    & 29.2 $\pm$ 0.1   & 457 $\pm$ 33    & 18.8$\pm$1.4$\pm$2.1    \\
    $\phi K^0   \bar{K}^{0}$                  & 6.2 $\pm$ 0.02   & 108 $\pm$ 11    & 21.0$\pm$2.1$\pm$2.3    \\
    $\phi K^0   \bar{K}^{*0} + \textrm{c.c.}$ & 4.0 $\pm$ 0.03   & 115 $\pm$ 17    & 34.6$\pm$5.1$\pm$2.1    \\
    $\phi K^+   K^-$                          & 25.2 $\pm$ 0.06  & 1067 $\pm$ 41   & 50.8$\pm$2.0$\pm$4.3    \\
    $\phi K^+   K^{*-}$                       & 17.7 $\pm$ 0.05  & 322 $\pm$ 27    & 21.8$\pm$1.8$\pm$1.4    \\
    $\phi K^-   K^{*+}$                       & 17.9 $\pm$ 0.05  & 333 $\pm$ 27    & 22.3$\pm$1.8$\pm$1.5    \\
    \hline
    \hline
    \end{tabular}
    \label{tab:xs_fit_res}
\end{table}
The cross sections of $e^+e^-\to\phi+\text{anything}$ and $e^+e^-\to\phi+X(s\bar{s})$ are also summarized in Table~\ref{tab:xs_fit_res}, where the first uncertainty is statistical and the second is systematic, as discussed in the following part.

In the calculation of $R_{s\bar{s}}$ via Eq.~\ref{eq:Rss},
the values of $f_{s\bar{s}}$ for $e^+e^-\to\phi\eta$ and $e^+e^-\to\phi\eta^{\prime}$ are set to 0.607$\pm$0.006 and 0.794$\pm$0.004~\cite{Balitsky:2015kwa}, respectively. 
For the $f_0(980)$, $f_2(1270)$, $f_0(1500)$, $f_2^{\prime}(1525)$, and other possible $f$ states, we assume that their $s\bar{s}$ components decay entirely into a pair of kaons.
Meanwhile, the $f_{s\bar{s}}$ values are taken to be unity for other $X(s\bar{s})$ processes.
In particular, we concentrate on the QCD-related contribution to $R_{s\bar{s}}$,
which requires subtracting the ISR processes, dominantly originating from the $e^+e^-\to\phi\gamma_\text{ISR}$ process.
A simultaneous fit of the number of photons and $\cos\theta_{\gamma}$, where $\theta_{\gamma}$ is the polar angle in the lab system, confirms that all events with $M_{\textrm{recoil}}(\phi) < 0.4~\text{GeV}/c^{2}$ stem from the $e^+e^-\to\phi\gamma_\text{ISR}$ process; the corresponding yield is extracted to be $1252\pm25_{\textrm{stat.}}$.
Additionally, $\epsilon(e^+e^-\to\phi+\textrm{anything})$ after subtracting the $e^+e^-\to\phi\gamma_\text{ISR}$ process is estimated to be $(26.6\pm0.4_{\textrm{stat.}})\%$ using the same reweighting method.
 The value of $R_{s\bar{s}}$ is determined to be $(40.4\pm1.7_{\textrm{stat.}})\%$ using the yields and efficiencies listed in Table~\ref{tab:xs_fit_res}.

\begin{table*}[!htbp]
     \centering
     \caption{The summary of systematic uncertainties of $\sigma/R_{s\bar{s}}$, given in percent. The ``-'' indicates that the corresponding source contributes no systematic uncertainty.}
     \begin{tabular}{llllllllll}
     \hline
     \hline
     Process                                               & Tracking & PID      & $K_{S}^0$ & $\phi$   & Bin width & Efficiency & Background & $\mathcal{LB}$ & Total     \\ \hline
     $e^+e^-\rightarrow\phi + \rm anything$                & 2.0/-    & 2.0/-    & -         & 1.7/1.67 & -         & 0.6/0.63   & 1.5/1.50   & 1.3/-          & 3.9/2.33  \\
     $e^+e^-\rightarrow\phi + \eta$                        & 2.0/-    & 2.0/-    & -         & -        & 1.0/0.12  & -          & 2.7/0.33   & 1.3/-          & 4.2/0.35  \\
     $e^+e^-\rightarrow\phi + \eta^{\prime}$               & 2.0/-    & 2.0/-    & -         & -        & 1.0/0.08  & -          & 10.8/0.85  & 1.3/-          & 11.3/0.85 \\
     $e^+e^-\rightarrow\phi K^0 \bar{K^{0}}$               & 2.0/-    & 2.0/-    & 1.5/0.17  & -        & 1.0/0.11  & 9.8/1.09   & 3.9/0.43   & 1.3/-          & 11.1/1.19 \\
     $e^+e^-\rightarrow\phi K^0 K^{*0}+\textrm{c.c.}$      & 2.0/-    & 2.0/-    & 1.5/0.27  & -        & 1.0/0.18  & 1.8/0.33   & 4.6/0.84   & 1.3/-          & 6.1/0.96  \\
     $e^+e^-\rightarrow\phi K^+ K^-$                       & 3.0/0.27 & 3.0/0.27 & -         & -        & 1.0/0.27  & 7.0/1.88   & 0.4/0.10   & 1.3/-          & 8.4/1.94  \\
     $e^+e^-\rightarrow\phi K^+ K^{*-}$                    & 3.0/0.12 & 3.0/0.12 & -         & -        & 1.0/0.12  & 2.3/0.27   & 4.1/0.47   & 1.3/-          & 6.5/0.58  \\
     $e^+e^-\rightarrow\phi K^- K^{*+}$                    & 3.0/0.12 & 3.0/0.12 & -         & -        & 1.0/0.12  & 3.5/0.41   & 3.6/0.42   & 1.3/-          & 6.8/0.62  \\
     ISR                                                   & -        & -        & -         & -        & -         & -          & -          & -              & -/1.02    \\
     \hline
     \hline
     \end{tabular}
     \label{tab:sys_sum}
 \end{table*}

Systematic uncertainties on the cross section measurements and $R_{s\bar{s}}$ arise from charged track detection, PID, $K_S^0$ reconstruction, $\phi$ reconstruction, $\phi$ fit, bin width selection, selection efficiency, and background description.
For $R_{s\bar{s}}$, an additional ISR correction contributes to its systematic uncertainty, whereas for the cross section measurements, extra systematic uncertainties stem from luminosity determination and the quoted branching fraction $\mathcal{B}(\phi\to K^+K^-)$.
The systematic uncertainties related to the two charged tracks originating from the $\phi$ meson are canceled in the $R_{s\bar{s}}$ determination, including track detection and PID, which are assigned as 1\% per track~\cite{BESIII:2016tbd}.
The systematic uncertainties due to $K_S^0$ reconstruction are determined to be $1.5\%$~\cite{BESIII:2015jmz} for each $K_S^0$ using the control samples of $J/\psi\rightarrow K^{*\pm}K^{\mp}$ and $J/\psi\rightarrow\phi K_S^0 K^{\pm}\pi^{\mp}$.
The systematic uncertainties associated with $\phi$ reconstruction and the mass fit are estimated by using a candidate selection algorithm that selects only the $\phi$ candidate with the smallest $|M_{K^+K^-}-m_{\phi}|$ and by varying the $\phi$ line shape in the fit, respectively, where $m_{\phi}$ is the nominal $\phi$ mass~\cite{ParticleDataGroup:2022pth}.
The systematic uncertainties associated with the bin width of the recoil mass distributions are estimated by varying the bin width for each channel and taking the relative change in the fitted signal yield as the uncertainty.
The systematic uncertainties originating from the background description are evaluated by varying the order of the polynomial,
the $f_0(980)$ line shape in the determination of the $e^+e^-\to\phi\eta^{\prime}$ process, and the definition of the sideband regions for background subtraction.
The uncertainties related to selection efficiency are treated as the difference between the nominal MC samples and the MC samples incorporating intermediate states, such as $f_0(980)$, $f_2(1270)$, $f_0(1500)$, $f_2^{\prime}(1525)$, and $K^*(1410)$.
The combined systematic uncertainty from the luminosity~\cite{BESIII:2021wib} and $\mathcal{B} (\phi\to K^+K^-)$~\cite{ParticleDataGroup:2022pth} is assigned as $1.3\%$ for the cross section measurements.
The uncertainty due to ISR correction is determined as the change in $R_{s\bar{s}}$ obtained by applying ISR correction factors for $e^+e^-\to\phi\eta/\eta'$~\cite{BESIII:2021bjn, BESIII:2020gnc} and the other processes~\cite{BESIII:2019ebn}.
Assuming that all these sources are independent, the total systematic uncertainties on the cross sections and $R_{s\bar{s}}$ are taken as the quadratic sum of the individual contributions.
The systematic uncertainties on the cross section measurements and $R_{s\bar{s}}$ are summarized in Table~\ref{tab:sys_sum}.
The cross sections are listed in Table~\ref{tab:xs_fit_res}, and $R_{s\bar{s}}$ is determined to be $(40.4\pm1.7_{\rm stat.}\pm1.5_{\rm syst.})\%$.

In summary, double-$s\bar{s}$ production is observed in electron-positron annihilation through a systematic measurement of $e^+e^-\to\phi + X(s\bar{s})$ using the $e^+e^-$ collision data at $\sqrt{s}=3.08$~GeV.
In addition, the cross section for $e^+e^-\rightarrow\phi + \textrm{anything}$ is measured for the first time,
offering an initial investigation into single inclusive $s\bar{s}$ hadron production in electron-positron annihilation.
The hadronization of $s\bar{s}$ pairs is investigated by measuring the cross sections of multiple processes: $e^+e^-\rightarrow\phi\eta\textrm{,}~\phi\eta'\textrm{,}~\phi K^+K^-\textrm{,}~\phi K^+K^{*-}\textrm{,}~\phi K^-K^{*+}\textrm{,}~\phi K^0\bar{K}^{0}\textrm{,}~\phi K^{0}\bar{K}^{*0}+\textrm{c.c.}$.
The cross sections for $e^+e^-\rightarrow\phi\eta\textrm{,}~\phi\eta'\textrm{,}~\phi K^+K^-$ confirm previous results~\cite{BESIII:2021bjn,BESIII:2020gnc,BESIII:2019ebn},
while the cross sections for $e^+e^-\rightarrow~\phi K^+K^{*-}\textrm{,}~\phi K^-K^{*+}\textrm{,}~\phi K^0\bar{K}^{0}\textrm{,}~\phi K^{0}\bar{K}^{*0}+\textrm{c.c.}$ are reported for the first time.

Furthermore, $R_{s\bar{s}}$ is measured to be $(40.4\pm2.3)\%$,
indicating that double-$s\bar{s}$ production is dominated by QCD.
Although processes with higher multiplicity, such as $e^+e^-\rightarrow\phi K^+K^- \pi^0$, are not included, the large value of $R_{s\bar{s}}$ confirms the phenomena observed in double-$c\bar{c}$ production~\cite{ref:Belle_2002, ref:Belle_2004abn, Belle:2009bxr, ref:BaBar_2005}.
In the context of the non-perturbative QCD effect, $R_{s\bar{s}}$ also challenges the existing theoretical framework developed to describe $R_{c\bar{c}}$~\cite{Braaten:2002fi, Bodwin:2002fk, Liu:2002wq, Hagiwara:2003cw, Liu:2004ga, Ma:2008gq}, thereby motivating a more complete theoretical understanding of hadronization and QCD.
Together with $R_{c\bar{c}}$, the observed $R_{s\bar{s}}$ provides a unique probe of potential underlying features of QCD and possible flavor-dependent mechanisms in hadronization.

\acknowledgments

The BESIII Collaboration thanks the staff of BEPCII (https://cstr.cn/31109.02.BEPC) and the IHEP computing center for their strong support. This work is supported in part by National Key R\&D Program of China under Contracts Nos. 2023YFA1606000, 2023YFA1606704, 2025YFA1613900; National Natural Science Foundation of China (NSFC) under Contracts Nos. 11635010, 11935015, 11935016, 11935018, 12025502, 12035009, 12035013, 12061131003, 12192260, 12192261, 12192262, 12192263, 12192264, 12192265, 12221005, 12225509, 12235017, 12342502, 12361141819, 12535005; the Chinese Academy of Sciences (CAS) Large-Scale Scientific Facility Program; the Strategic Priority Research Program of Chinese Academy of Sciences under Contract No. XDA0480600; CAS under Contract No. YSBR-101; 100 Talents Program of CAS; The Institute of Nuclear and Particle Physics (INPAC) and Shanghai Key Laboratory for Particle Physics and Cosmology; Agencia Nacional de Investigaci\'on y Desarrollo de Chile (ANID), Chile under Contract No. ANID CCTVal CIA250027; ERC under Contract No. 758462; German Research Foundation DFG under Contract No. FOR5327; Istituto Nazionale di Fisica Nucleare, Italy; Knut and Alice Wallenberg Foundation under Contracts Nos. 2021.0174, 2021.0299, 2023.0315; Ministry of Development of Turkey under Contract No. DPT2006K-120470; National Research Foundation of Korea under Contract No. RS-2026-25486791; National Science and Technology fund of Mongolia; Polish National Science Centre under Contract No. 2024/53/B/ST2/00975; STFC (United Kingdom); Swedish Research Council under Contract No. 2019.04595; U. S. Department of Energy under Contract No. DE-FG02-05ER41374.
This paper is also supported by the Innovation Fund of IHEP under Contract No. E55452U2, the Start-up Fund of IHEP under Contract No. E55156U1, and the National Natural Science Foundation of China under grant No. 12342502.
\newline


%

\end{document}